\documentstyle[twoside,fleqn,espcrc2]{article}

\newcommand{\NCS}{N_{\rm CS}}

\title{The Nonperturbative Broken Phase Sphaleron Rate}

\author{Guy D. Moore\address{Physics Department, McGill University,
	3600 University Street, Montreal, QC H3A 2T8 Canada}}
       
\begin{document}

\begin{abstract}
I present a technique for measuring the broken phase sphaleron rate
nonperturbatively.  There are three parts to the calculation:
determination of the probability distribution of Chern-Simons number
$\NCS$; measurement of $\langle | d\NCS/dt | \rangle_{N_{\rm
CS}=1/2}$, the mean rate of 
change of $\NCS$ at the barrier; and measurement of the
``dynamical prefactor,'' the fraction of barrier crossings which result
in a permanent integer change in $\NCS$.
\end{abstract}

\maketitle

\section{Introduction}

The ``sphaleron rate,'' the spacetime diffusion
constant for Chern-Simons number $\NCS$, is a key quantity in the study
of electroweak baryogenesis, setting the efficiency of baryon number
violation.  It is also a very appropriate topic at this
conference, since it is nothing but the topological susceptibility of the
electroweak plasma, only in {\it Minkowski}, time;
\begin{equation}
\Gamma \equiv \lim_{t \rightarrow \infty} \frac{ \langle ( \NCS(t) - 
	\NCS(0) )^2 \rangle }{Vt} \, .
\end{equation}

$\Gamma$ can be determined reliably by perturbation theory in any regime
where $g \phi(T)/T \gg \alpha_w$, where $\phi(T)$ is the Higgs
condensate.  However, this is precisely the regime where it is so small
as to be of no physical consequence.  We really want to know it in the
symmetric phase at $T_c$ the phase transition temperature, to determine
how many baryons can be produced; and in the broken phase at and
immediately below $T_c$ to determine whether they are subsequently
erased or are preserved until the present epoch.

Recently there has been significant progress in the symmetric phase
rate.  It has been shown analytically \cite{ASY} and verified
numerically \cite{particles} that ``hard thermal loop'' radiative
corrections are important, and the rate is known with reasonable
accuracy, though it has recently been demonstrated that there are
logarithmic corrections \cite{Bodek_log} which are not well under
control.  

But the real time method cannot measure the broken phase rate; the rate
is so low that no events will occur in any reasonable amount of time.
Since we cannot extend the perturbative calculation beyond 1 loop, 
we will instead go about a nonperturbative measurement.
We can use the dimensional reduction approximation and we will
relentlessly; we also take the dynamics to be classical, regulated to
give a physically reasonable cutoff to the hard thermal loops.
For a much more detailed discussion of the work presented here, see
\cite{broken_long}.

\section{Defining and measuring $\NCS$}

\begin{figure}[htb]
\vspace{0.6in}
\includegraphics{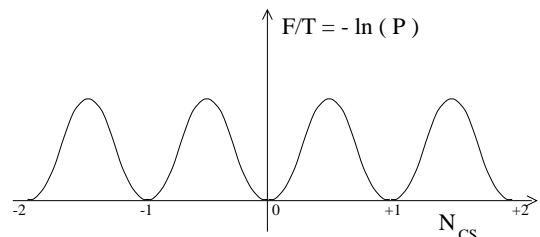}
\caption{Cartoon of the free energy dependence on $\NCS$}
\vspace{-0.3cm}
\end{figure}

Here is the cartoon picture of why baryon number violation is slow in
the broken phase.  Because the Higgs condensate makes being half 
way between winding numbers energetically expensive, there is a free
energy barrier to being at $\NCS=1/2$.  Our method of determining the
rate is to find the probability distribution of $\NCS$, and to multiply
the chance to be in a narrow range about $\NCS=1/2$ by the mean crossing
time of that range, set by $\langle | d\NCS/dt | \rangle$.  Then we need
to determine what fraction of crossings are permanent.

In the continuum, the meaning of $\NCS$, modulo 1, is
the integral of $F \tilde{F}$,
\begin{equation}
\NCS \equiv \int d\tau \left( \frac{g^2}{8 \pi^2} \int d^3 x
	B_i^a \left[ D_\tau , A_i \right]^a \right) \, .
\end{equation}
Here $\tau$ parameterizes a path through the space of 3-D configurations.
The constant of integration is set by defining $\NCS=0$ in
vacuum.  This is consistent since the above integral around any closed
loop is the second Chern class of the loop, which is an integer.
However, to make a nonperturbative study we need to find a lattice
implementation for the above, which is touchy since no lattice
definition of $F \tilde{F}$ is a total derivative.  The best thing to do
is to use the above definition for a unique specific path,
the cooling path, defined by
\begin{equation}
\frac{dU}{d\tau} = (D^\alpha U) ( D^\alpha H_{\rm YM} ) \, ,
\end{equation}
with $H_{\rm YM}$ the Yang-Mills part of the 3-D Hamiltonian.
This definition of $\NCS$ is unique because the cooling path is; it is
also minimally exposed to lattice problems since the cooling process
very quickly eliminates UV fluctuations, which are responsible for any
poor behavior in $F \tilde{F}$.

There are three problems we must take care of, though; a UV noise
problem, a numerical cost problem, and the problem of turning a
probability distribution for $\NCS$ into a rate.

\section{The first two problems}

The first problem is that $\NCS$ actually is not a very good measurable
to be using.  The problem is that it contains a lot of UV noise not
related to topology.  Our discussion will follow that of
Ambj{\o}rn and Krasnitz \cite{AmbKras}.  In the UV the 3-D classical thermal
theory we are studying behaves at leading order like 3 copies of the
abelian theory, in which
\begin{equation}
\NCS = \frac{g^2}{32 \pi^2} \int d^3 x \epsilon_{ijk} 
	F_{ij}^a A_k^a \, .
\end{equation}
The mean square value at leading order in perturbation theory is then
\begin{equation}
\langle \NCS^2 \rangle = \frac{g^4 T^2 V}{64 \pi^4} \int
	\frac{d^3 p}{(2 \pi)^3} \frac{1}{p^2} \, ,
\end{equation}
which is linearly UV divergent.  The size of the divergence is extensive
in volume and will be cut off by the lattice spacing, leading to a $1/a$
behavior.

The solution is not to measure $\NCS$ itself, but $\NCS$ {\em after}
some preliminary cooling has been performed;
\begin{equation}
N \equiv \int_{\tau_0}^{\infty} d \tau \left( \frac{g^2}{8 \pi^2} 
	\int d^3 x B_i^a \left[ D_\tau , A_i \right]^a \right) \, .
\end{equation}
To illustrate that this does the trick, we show $\NCS$ and $N$ for
a period of Hamiltonian evolution in the broken phase, and $N$ for a
sphaleron transition:  

\begin{figure}[htb]
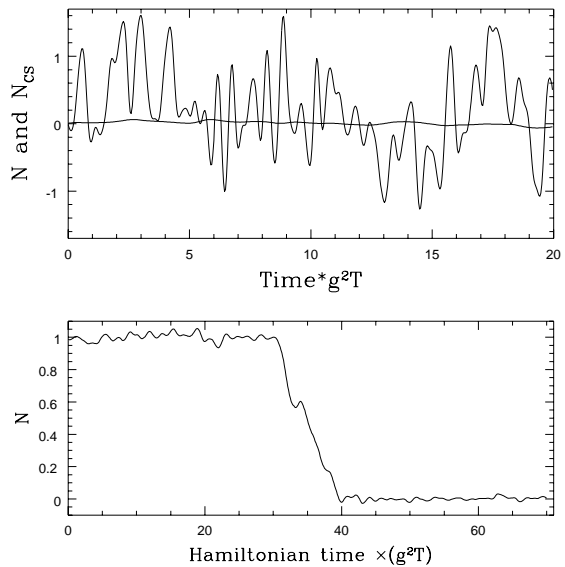

\vspace{1.05in}
\includegraphics{NvsNCS.epsi}
\vspace{1.45in}
\includegraphics{no_particles.epsi}
\caption{Above:  $\NCS$ (oscillating like mad) and $N$ (always near
zero) for a Hamiltonian trajectory in the broken phase.  Below:  $N$
during a sphaleron.}
\vspace{-0.5cm}
\end{figure}

\noindent
This graphically illustrates that using $N$ cuts out irrelevant UV
noise, while still capturing interesting topological information.

There is still a problem with using $N$, the numerical cost
involved in performing the cooling, which we must do 
thousands of times to get an accurate Monte-Carlo study of the
probability distribution.  We finesse this problem by observing that,
since cooling eliminates the UV excitations, it quickly renders the
lattice fields smooth enough that we lose no information by blocking by
a factor of 2.  When we use a classically improved Hamiltonian and
definition of $F \tilde{F}$ after the blocking, there is 
almost no difference between the blocked and unblocked measurement of
$N$, provided we cool by
at least $\tau > 1.25 a^2$ before blocking.  For lattices larger than
about $28^3$ we can double block.

\section{Turning a probability into a rate}

We can now find the ($\tau_0$ dependent) probability distribution of $N$
by conventional multicanonical means.  To turn this into a
rate, we first measure $\langle |dN/dt| \rangle$ (sampled over
$N=1/2$ configurations) by making a canonically
distributed sample of $N=1/2$ configurations, generating momenta from the
canonical ensemble, and evolving for a very short Hamiltonian time,
measuring $N$ before and after.  The mean rate of $N=1/2$ crossings per
4-volume is then 
\begin{equation}
\frac{\Gamma}{\rm prefactor} = \frac{P(|N-0.5| < \epsilon)}{2 \epsilon}
	\times \frac{\langle | dN/dt | \rangle}{V} \, .
\end{equation}

To measure the dynamical prefactor, which is the fraction of crossings
which lead to permanent integer change in $\NCS$, we use the same
$N=1/2$ sample but continue the Hamiltonian time evolution, both forward
and backward in time, until the system settles about a vacuum; then we
count how many crossings and how many permanent $N$ changes occur.  The
fraction of crossings leading to permanent $N$ change is
\begin{equation}
\frac{1}{N_{\rm samp}} \sum
	\frac{1}{{\rm \# \; crossings}} \times \left\{
	\begin{array}{cc} 1 & {N_{\rm fin} \neq N_{\rm init}} \\
		          0 & {N_{\rm fin}=N_{\rm init}} \\ 
	\end{array} \right.  .
\end{equation}
For very large Debye mass squared $m_D^2$ there are multiple crossings,
but for realistic $m_D^2$ the prefactor is about 0.4.

\section{Results}

\begin{figure}[htb]
\vspace{2.25in}
\includegraphics{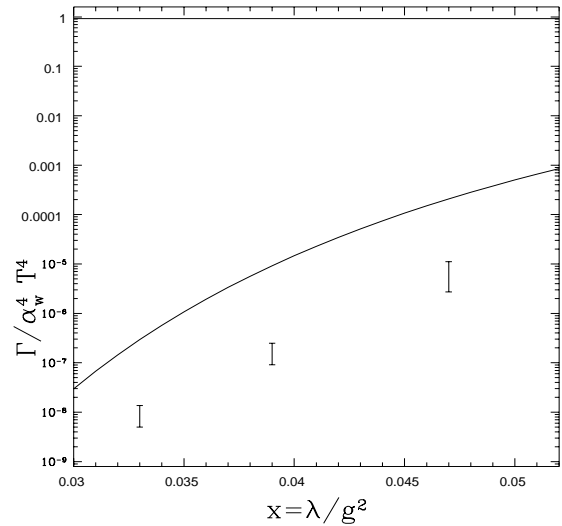}
\caption{\label{lastfig}Nonperturbative value of $\Gamma(T_c)$ (points
with bars) versus
perturbative estimate (curve) and the symmetric phase rate (curve at
top), as a function of $x \equiv \lambda / g^2$.}
\vspace{-0.15in}
\end{figure}

The final
results for $\Gamma$ are compared to perturbation theory (two loop
effective potential but tree kinetic terms) in Figure
\ref{lastfig}.  Baryon number erasure after the phase transition is
prevented if $\Gamma < 10^{-7} \alpha_W^4 T^4$, which occurs if 
$x < 0.037$.

\begin{figure}[htb]
\vspace{1.30in}
\includegraphics{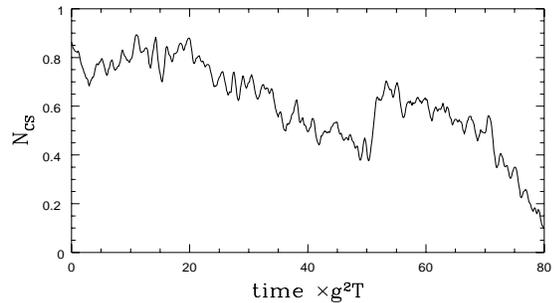}
\caption{Multiple crossings when $m_D^2 = 43 g^4 T^2$, a very large hard
thermal loop strength.}
\end{figure}

\end{document}